\documentclass[conference]{IEEEtran}

\usepackage{braket}
\usepackage{amsmath}
\usepackage{amssymb}
\usepackage{graphicx}
\usepackage{cite}
\usepackage{epstopdf}
\usepackage{mathtools}
\usepackage{amsthm}
\usepackage{url}

\newcommand{\tr}{\text{Tr}}
\newcommand{\ketbra}[1]{\ket{#1}\bra{#1}}

\title{Improving QKD for Entangled States with Low Squeezing via Non-Gaussian Operations}

\author{
	\IEEEauthorblockN{Eduardo Villase\~nor and Robert Malaney}\\
	\IEEEauthorblockA{School of Electrical Engineering  \& Telecommunications,\\
		The University of New South Wales, Sydney, NSW 2052, Australia.}\\
}

\begin{document}

\maketitle

\IEEEpeerreviewmaketitle

\begin{abstract}
In this work we focus on evaluating the effectiveness of two non-Gaussian operations, photon subtraction (PS) and quantum scissors (QS) in terms of Continuous Variable (CV)-Quantum Key Distribution (QKD) over lossy channels.  Each operation is analysed in two scenarios, one with the operation applied transmitter-side to a Two-Mode Squeezed Vacuum (TMSV) state and a second with the operation applied to the TMSV state receiver-side. We numerically evaluate the entanglement and calculate the  QKD key rates produced in all four possible scenarios. Our results show that for a fixed value of initial squeezing in the TMSV state, the states produced by the non-Gaussian operations are more robust to loss, being capable of generating higher key rates for a given loss. More specifically, we find that for values of initial TMSV squeezing below 1.5dB the highest key rates are obtained by means of transmitter-QS. On the other hand, for squeezing above 1.5dB we find that receiver-PS produces higher key rates. Our results will be important for future CV-QKD implementations over free-space channels, such as the Earth-satellite channel.
\end{abstract}

\section{Introduction}
There is great interest in the use of continuous variable (CV) photonic states for quantum information processing\cite{braustein2005QIwithCV,PhysRevLett.82.1784}. In many CV-based protocols, CV
 entanglement  plays a pivotal role \cite{entanglement}. However,
perfectly entangled bipartite CV states, represented by two-mode squeezed vacuum (TMSV) states with infinite squeezing are non-physical.
Nonetheless, finite-squeezed TMSV states (producing non-perfectly correlated measurement outcomes) are still valuable resources for CV quantum communication protocols, and can be readily created experimentally \cite{PhysRevLett.59.2555}.

In current experiments TMSV squeezing  up to 16dB is achievable, albeit difficult to produce \cite{15dBSqueeze,16dBSqueeze}.
However, states produced up to 5dB are easily achievable in the laboratory \cite{PhysRevA.73.063804,6dB}.
Besides practical limitations in squeezing, imperfections in the devices used to create the TMSV states and the unavoidable interaction of the state with the environment (photonic loss) both introduce unwanted noise, eventually lowering the amount of entanglement that can be utilised.
To correct for these effects and to increase the amount of entanglement shared between communicating parties we require some form of entanglement distillation \cite{DistillationOriginal}. Such distillation can be considered a special form of quantum error correction \cite{Purification}.

However, when dealing with CV quantum states certain restrictions need to be considered when we investigate distillation. A  significant result in this regard is the no-go theorem for Gaussian entanglement distillation \cite{nogoDistillation}, which states that it is impossible to distil Gaussian quantum states by means of Gaussian operations and Gaussian measurements only.
This is of great importance as many of the errors that arise in naturally occurring channels are Gaussian \cite{Holevo2007}.
To remedy this situation, non-Gaussian operations are often turned to as a means of delivering entanglement distillation within quantum information protocols \cite{CVdistillation}.

For Gaussian states, CV quantum key distribution (QKD) \cite{PhysRevA.87.022308} - arguably the most important quantum information protocol - has been well investigated both experimentally and theoretically \cite{GaussianQuantumInformation}.
Nonetheless, CV-QKD delivered by means of non-Gaussian states has shown great potential,  largely due to the fact that for a given  TMSV state non-Gaussian operations acting on it can probabilistically increase the entanglement \cite{EntanglementPS, DistillationPS}.
Previous works along these lines have analysed the effectiveness of non-Gaussian operations such as photon subtraction
(PS) in increasing entanglement \cite{PSoriginal, EntanglementPS,CVdistillation}. In addition, the subsequent improvement in
the key rates of QKD schemes via PS has been investigated, e.g. \cite{MingjianPS, MingjianPAPS}.
Another non-Gaussian operation known as quantum scissors (QS) \cite{original_scissors} has also surfaced as a prominent tool for entanglement distillation of CV states \cite{scissors2, scissors}. QS has also been proven to increase the key rates in QKD protocols under specific conditions \cite{ScissorsQKD}.

However, what is currently lacking in the present literature is a systematic comparison, under lossy channel conditions, between the effectiveness of PS vs QS in terms of the resultant QKD rates, and whether transmitter-side or receiver-side operations are best in this regard. Such a systematic comparison is the contribution of this present work.

\section{System description}
We consider two parties, Alice and Bob, who are connected via a lossy channel.
As shown in Fig.\ref{fig:channel}(a), Alice is capable of generating an entangled TMSV state from which she will communicate one mode to Bob via the channel (whilst retaining the other mode).
Before transmitting a mode to Bob, Alice can choose to apply either QS or PS to it; we call this scenario the transmitter-operation.
Conversely, in the receiver-operation scenario Alice can choose to transmit a mode without acting on it operationally,  and on receipt Bob applies an operation.
Alternatively, neither Alice or Bob can apply any operation and simply use the TMSV state in the protocol.
Thus, we have five different scenarios: TMSV, receiver-PS, transmitter-PS, receiver-QS and transmitter-QS.

When dealing with Gaussian states and Gaussian operations there exists a well developed theoretical formalism based on the first two statistical moments of the Gaussian states \cite{GaussianQuantumInformation}.
However, for non-Gaussian states the leading two statistical moments are not enough to fully represent the states, and we are forced into more complex calculations based on, say, the density matrix formalism. In this work we adopt this latter formalism and numerically simulate the evolution of the quantum states in order to determine the effects of non-Gaussian operations and photon loss.

For a single mode the Hilbert space is represented in the Fock number basis $\{ \ket{n}_{n=0}^\infty \}$. In the calculations presented we limit the Hilbert space of each individual mode to the first 20 number states of the Fock basis. This limited Hilbert space is still large enough to include all the most important coefficients in the Fock basis for the states we  consider.

\subsection{Evolution of the quantum states}
Here we describe the evolution of the bipartite quantum state as one mode is transmitted though the channel and a non-Gaussian operation is applied to it in the receiver-side case. For the case where the operation is applied transmitter-side the same description applies with the difference that the non-Gaussian operations act before the channel.

We begin with a TMSV state previously prepared by Alice in the modes $A$ and $B$. The TMVS with real squeezing parameter $r_{AB}$, is represented in the Fock number basis as:
\begin{align}
\ket{\psi}_{AB} = \frac{1}{\cosh(r_{AB})}\sum_{n=0}^\infty(-\tanh(r_{AB}))^n \ket{n n}.
\end{align}
In the results presented below we will measure the squeezing in dB as $r_{AB}^{dB}= 10\log_{10}(\exp(2r_{AB}))$.


An important operation in optics is the beam splitter (BS), which we will use as component of the non-Gaussian operations, and as a means to model the lossy channel. In general, the evolution of two modes $i$ and $j$ interacting in a BS of transmissivity $t$ is defined by the operator:
\begin{align}
\hat{U}^{ij}_t := \exp[\arccos{(\sqrt{t})}(\hat{a}_i^\dagger \hat{a}_j - \hat{a}_i \hat{a}_j^\dagger)],
\label{eq:bs}
\end{align}
with $\hat{a}_i$ and $\hat{a}_j$ the annihilation operators of modes $i$ and $j$, respectively.

As shown in Fig.\ref{fig:channel}(b), when Alice transmits mode $B$ to Bob we model the loss in the channel by means of a BS of transmissivity $\eta$ in which mode $B$ interacts with a vacuum in mode $E'$. The evolution of the state is obtained by applying the operator $\hat{U}^{BE'}_\eta$, following eq. \ref{eq:bs}.
After the interaction in the BS the environment mode $E'$ is traced out, resulting in a mixed state.
Therefore, the output state after the effect of the lossy channel $\mathcal{M}$ is
\begin{align}
\rho'_{AB} = \tr_{E'}( \hat{U}^{BE'}_\eta (\ket{\psi}\bra{\psi}_{AB} \otimes \ketbra{0}_{E'} ){\hat{U}^{BE'}_\eta}{}^ \dagger).
\end{align}
Throughout this work we measure the attenuation caused by photon loss in dB using $-10\log_{10}(\eta)$.
Next, we describe the two non-Gaussian operations that are the focus of this work.

\subsection{Noiseless linear amplification via quantum scissors}
The QS operation considers an input mode and two auxiliary modes, a vacuum and a single photon.
As shown on Fig.\ref{fig:channel}(c), two auxiliary modes $C'$ and $C$ interact via a BS with transmissivity $\kappa_{QS}$, and input mode $B$ interacts with mode $C$ in a 50/50 BS resulting in the state:
\begin{align}
\rho''_{AB}=\hat{U}^{BC}_{1/2}\hat{U}^{CC'}_{{\kappa_{QS}}}( \rho'_{AB} \otimes  \ketbra{10}_{CC'} ) \hat{U}^{CC'}_{\kappa_{QS}}{}^ \dagger \hat{U}^{BC}_{1/2}{}^ \dagger.
\end{align}
Thereafter, modes $B$ and $C$ are measured by a pair of single-photon detectors, $D_B$ and $D_C$, respectively.
If detector $D_B$ successfully detects a photon, while $D_C$ does not (or vice-versa), then the protocol is successful. The measurement is represented by the action of the projector $\Pi_{10} = \Pi^B_1 \otimes \Pi^C_0$ (or $\Pi_{01} = \Pi^B_0 \otimes \Pi^C_1$), with $\Pi_{b \in \{0, 1\}} = \ketbra{b}$, which causes the state to collapse to
\begin{align}
\rho_{\text{out}} = \frac{\Pi_{10} \rho''_{AB} \Pi_{10}^\dagger}{\tr(\Pi_{10} \rho''_{AB})}.
\end{align}

The QS operation is a probabilistic heralded operation with a probability of success  $P_{s} = \tr((\Pi_{01} + \Pi_{10}) \rho''_{AB})$.
The effect of the operation is a resulting state truncated over the Fock space.
Explicitly, given any input state  $\ket{\phi}=\alpha_0\ket{0} + \alpha_1\ket{1} + ...$ the QS operator $\mathcal{QS}$ acts in the following manner
\begin{align}
\mathcal{QS}(g, \alpha_0\ket{0} + \alpha_1\ket{1} + ...) = \Gamma (\alpha_0\ket{0} + g\alpha_1\ket{1}),
\end{align}
with $g$ the gain parameter set by $\kappa_{QS} =1/(1 + g^2)$ \cite{original_scissors}, and $\Gamma$ a normalization factor.
When the input state $\ket{\phi}$ is such that it mainly resides in the subspace onto which the QS truncate, then QS effectively produce noiseless linear amplification (NLA), namely $\mathcal{QS}(g, \ket{\phi})=\ket{g\phi}$.
That is,  for any input state $\ket{\phi}$, it is required that
\begin{align}
\ket{\phi} \approx \alpha_0\ket{0} + \alpha_1 \ket{1}
\label{eq:stateQS}
\end{align}
for QS to approximate NLA.
On the contrary, for states that do not satisfy eq. \ref{eq:stateQS} the QS operation will not be described by $\mathcal{QS}(g, \ket{\phi})=\ket{g\phi}$.

As a side note, a more complex scheme involving multiple QS has been introduced \cite{ScissorsNLA}. In this scheme, multiple QS operations are involved to implement NLA for any general input state.
We will restrict our analysis to an operation consisting of a single QS, since in the generalized NLA the probability of success of the entire operation drops exponentially with the number of QS used.

\subsection{Photon subtraction}
\begin{figure}
\centering
\includegraphics[width=.49\textwidth]{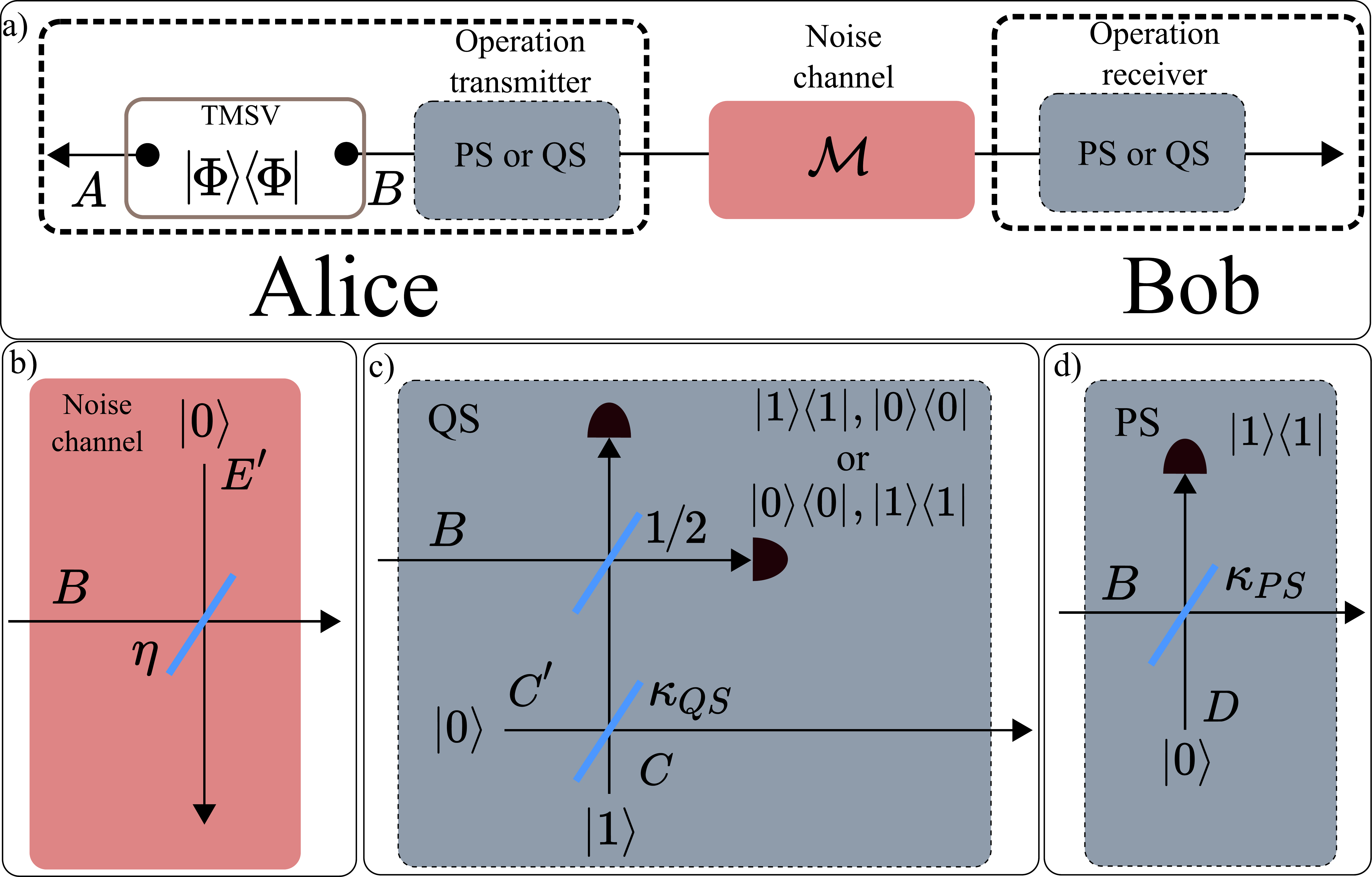}
\caption{a) A single mode of a TMSV pure state is transmitted from Alice to Bob over a lossy channel. b) We model the lossy channel by a vacuum interacting with the input mode in a BS.
c) The QS operation consists of an input mode, single photon injection, two BSs and single photon detectors.
d) In PS the input mode interacts with a vacuum state in a BS, afterwards if a single-photon is detected the operation was successful.
}
\label{fig:channel}
\end{figure}

The second non-Gaussian operation of interest is PS.
As shown in Fig.\ref{fig:channel}(d), PS over a single mode begins by mixing the input mode with a vacuum state in a BS of transmissivity $\kappa_{PS}$, resulting in
\begin{align}
\rho''_{AB}=\hat{U}^{BD}_{{\kappa_{PS}}}( \rho'_{AB} \otimes  \ketbra{0}_{D})  \hat{U}^{BD}_{\kappa_{PS}}{}^ \dagger.
\end{align}
Thereafter, a single-photon detection takes place, collapsing the state to
\begin{align}
\rho_{\text{out}} = \frac{\Pi_{1}^D \rho''_{AB} \Pi_{1}^D{}^\dagger}{P_{s}},
\end{align}
with the probability of success of the operation $P_{s} = \tr(\Pi_{1}^D \rho''_{AB})$.
Just like QS, PS is a heralded operation.
For simplicity, in both QS and PS we assume single-photon detector efficiencies of 100\% and that single photons can be generated on command.

\section{Calculating the key rate in CV-QKD}
\begin{figure}
\centering
\includegraphics[width=.47\textwidth]{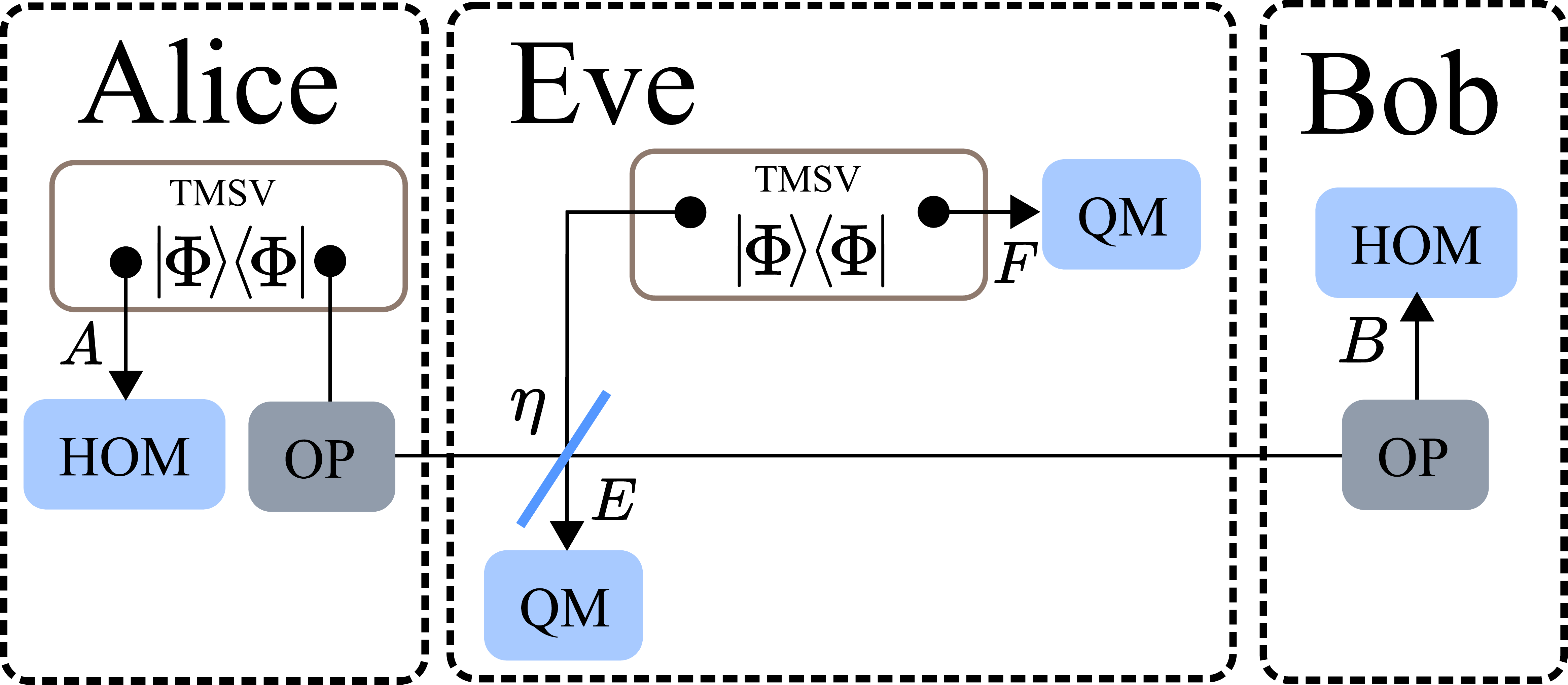}
\caption{The entanglement based CV-QKD protocol \cite{PhysRevA.87.022308}. Alice transmits the mode $B$ of a TMSV state to Bob. Separately, Eve prepares a TMSV state of her own, and sends mode $E$ through a BS that mimics the channel.
Afterwards, she can utilize any measurement on her modes to extract all the information possible.
Here OP = QS or PS, HOM = homodyne measurement and QM = quantum measurement.
}
\label{fig:QKD}
\end{figure}

We compute the secret key rate for the entanglement-based CV-QKD protocol described in \cite{PhysRevA.87.022308}.
As pictured in Fig.\ref{fig:QKD}, in our protocol Alice generates a TMSV state in modes $A$ and $B$ and transmits mode $B$ to Bob, with a non-Gaussian operation applied to $B$ (either transmitter-side or receiver-side).

To compute the key rate we follow the procedure presented in \cite{MingjianPS}.
Since we are considering that the states involved are non-Gaussian, the calculations of the key rates are more complex relative to those required to obtain key rates using Gaussian states.
Nonetheless, it has been proven that for any given covariance matrix associated with a non-Gaussian state, the key rate calculated from it corresponds to a lower bound \cite{ExtremalityGaussian}.
Thus, we will make use of this result emphasizing that the key rates presented here represent only a lower bound of the real values. For a complete discussion of how to calculate the key rate when non-Gaussian states are involved we refer the reader to \cite{MingjianPS, MingjianPAPS}.

During the QKD protocol we assume there is an eavesdropper, Eve, performing a collective attack \cite{GaussianAttack}.
In one form of a  collective attack we consider that Eve starts with a weakly squeezed TMSV state
$\ket{\psi}_{EF}$. We set the variance of $\ket{\psi}_{EF}$ to $1.002$ to model additional channel noise. Eve mimics the channel by using a BS of transmissivity $\eta$ in which the transmitted mode $B$ interacts with her mode $E$.
After the interaction in the BS, Eve keeps her modes $E$ and $F$, and is capable of performing any measurement on them or storing them indefinitely as required.
In this way Eve can obtain all the information lost during the transmission of mode $B$ through the channel.\footnote{This specific form of attack is not optimal for non-Gaussian states \cite{MingjianPS}, but the bound on key rate we derive remains true.}

When all the considerations above are accounted for, the key rate in the asymptotic limit of infinitely many uses of the channel is lower bounded as follows \cite{PhysRevA.87.022308}
\begin{align}
K \geq P_s(I_{AB} - \chi_{BE}),
\label{eq:key_rate}
\end{align}
with $P_s$ the probability of success of the used operation (QS or PS), $I_{AB}$ the mutual information between Alice and Bob, and $\chi_{BE}$ the Holevo information that Eve can extract from her measurements.
In the calculation of $I_{AB}$ we are assuming an efficiency of 1 in the classical post-processing. Additionally, we are considering only the successful applications of the non-Gaussian operations are used in our QKD protocol.

Now we compute the quantities $I_{AB}$ and $\chi_{BE}$.
We define the vector containing the quadrature operators $\hat{q}_i = \frac{\hat{a}_i + \hat{a}^\dagger_i}{\sqrt{2}}$ and $\hat{p}_i = \frac{\hat{a}_i - \hat{a}^\dagger_i}{i\sqrt{2}}$ ($\hbar$ = 1) of all the modes
\begin{align}
\hat{x} := (\hat{q}_A, \hat{p}_A,  ...., \hat{q}_F, \hat{p}_F).
\end{align}
The elements of the covariance matrix of the state are defined as
\begin{align}
V_{ij} := \langle\{ \Delta \hat{x}_i,  \Delta \hat{x}_j \} \rangle_{\rho_{\text{out}}},
\label{eq:covmax}
\end{align}
with $\Delta \hat{x}_i = \hat{x}_i - \langle \hat{x}_i \rangle$ and $\{ .\}$ the anti-commutation operation.
Using eq. \ref{eq:covmax} to calculate the covariance matrix entries for the modes $A$ and $B$ we obtain the following symmetric matrix
\begin{align}
M_{AB} =
\begin{bmatrix}
V_{AA} I & V_{AB} Z \\
V_{AB} Z & V_{BB} I
\end{bmatrix},
\end{align}
with $I$ the $2 \times 2$ identity matrix and $Z=diag(1,-1)$.
In this case the mutual information is given by the equation
\begin{align}
I_{AB} = \frac{1}{2}\log_2 \frac{V_{BB}}{V_{B|A}},
\end{align}
with the conditional variance $V_{B|A}=V_{BB} - \frac{{V_{AB}}^2}{V_{AA}}$.

For the Holevo information $\chi_{BE}$, we calculate it as
\begin{align}
\chi_{BE} = \sum_i g(\nu^{EF}_i) - \sum_i g(\nu^{EF|B}_i),
\end{align}
with
\begin{align*}
g(x) = \frac{x+1}{2}\log_2 \frac{x+1}{2} -  \frac{x-1}{2}\log_2 \frac{x-1}{2}.
\end{align*}
Here $\{\nu^{EF}_i\}$ and $\{\nu^{EF|B}_i\}$ correspond to the symplectic eigenvalues of the covariance matrices $M_{EF}$ and $M_{EF|B}$ respectively. The latter corresponding to the case when a homodyne measurement is made on the mode $B$,
\begin{align}
M_{EF|B} = M_{EF} -
\begin{bmatrix}
V_{EB} I \\
V_{FB} Z
\end{bmatrix}
\begin{bmatrix}
V_{BB}^{-1} & 0 \\
0 & 0
\end{bmatrix}
\begin{bmatrix}
V_{EB} I \\
V_{FB} Z
\end{bmatrix}^T,
\end{align}
with $V_{EB}$ and $V_{FB}$ the elements in the covariance matrix of the respective modes.

\section{Numerical results}
Now we present the results obtained in terms of the entanglement distilled and the QKD key rates produced.

\subsection{Entanglement properties}
To quantify entanglement we focus on the logarithmic negativity, which has been shown to be an entanglement monotone and an upper bound to the distillable entanglement, besides being remarkably easy to compute \cite{LogNeg}.
The logarithmic negativity quantifies how much a multipartite state is separable by measuring how much it fails to satisfy the positivity of the partial transpose with respect to its partitions.
In this case, computed as
\begin{align}
E_N(\hat{\rho}_{out})=\log||\hat{\rho}_{out}^{T_B}||_1.
\end{align}
However, for Gaussian states it can be computed from the symplectic eigenvalues $\{\nu_i\}$ of the respective covariance matrix as
\begin{align}
E_N(\hat{\rho}_{out}) = \sum_i F(\nu_i),
\end{align}
with $F(x)=-\log(x)$ for $x<1$ and $F(x)=0$ otherwise.

We compute $E_N$ for different values of initial squeezing in the TMSV state and for three fixed values of photon loss in the channel.
Here the transmissivities of the BSs in the non-Gaussian operations are optimal to maximize $E_N$, with the values $\kappa_{PS}=0.95$ and $\kappa_{QS}=0.05$.
We will fix these values throughout the rest of this work unless stated otherwise.

\begin{figure}
\centering
\includegraphics[width=.41\textwidth]{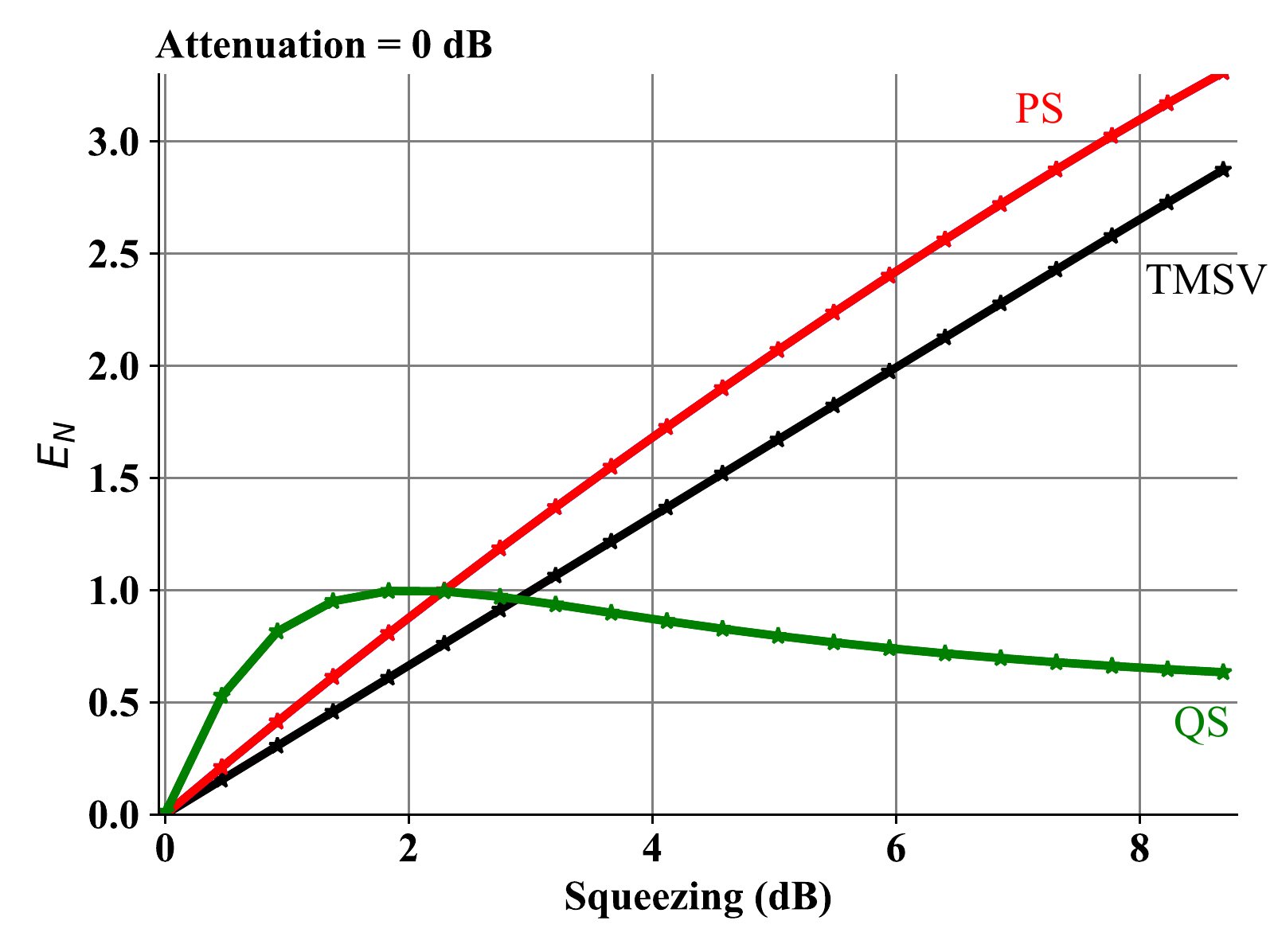}
\includegraphics[width=.41\textwidth]{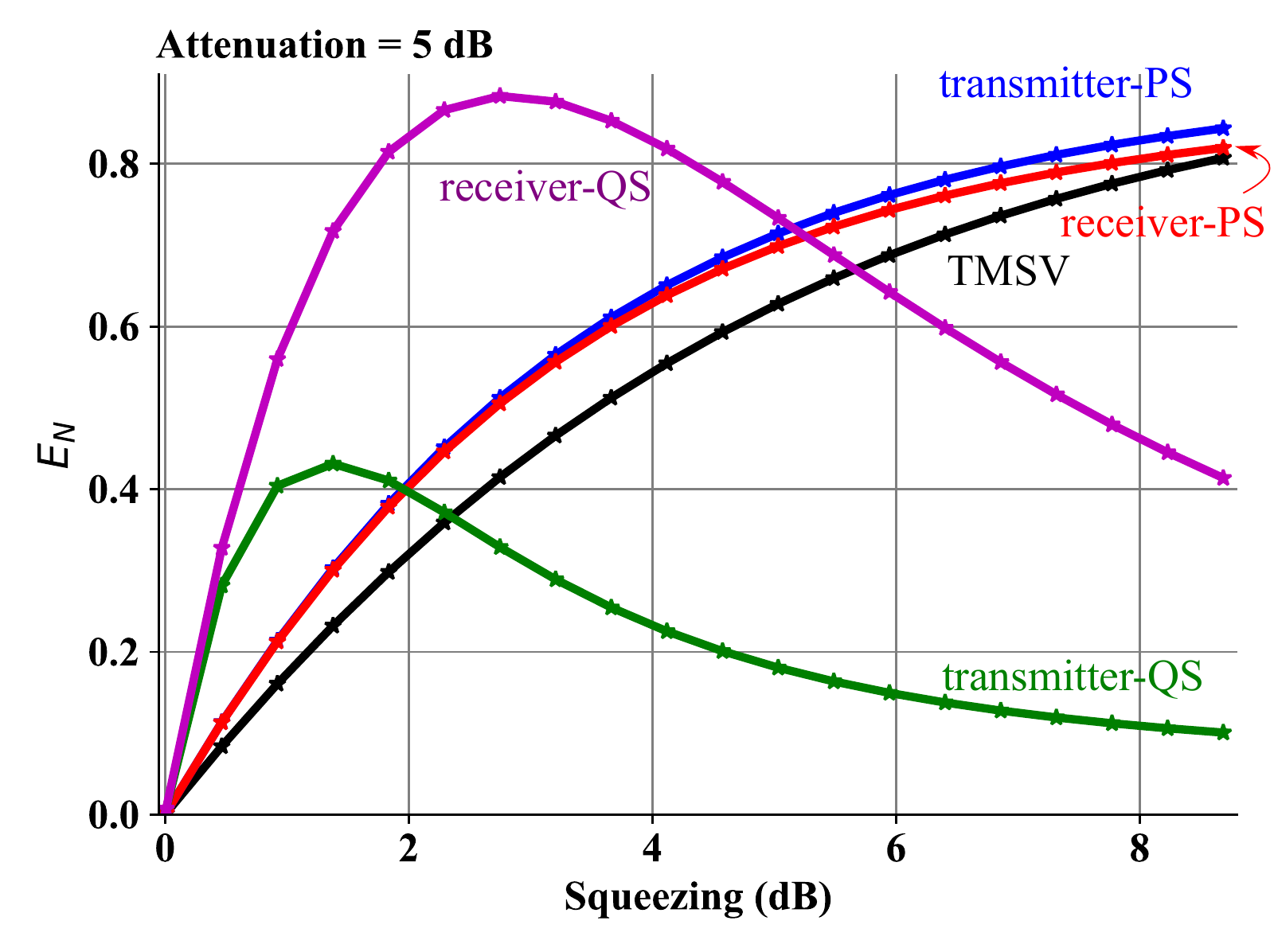}
\includegraphics[width=.41\textwidth]{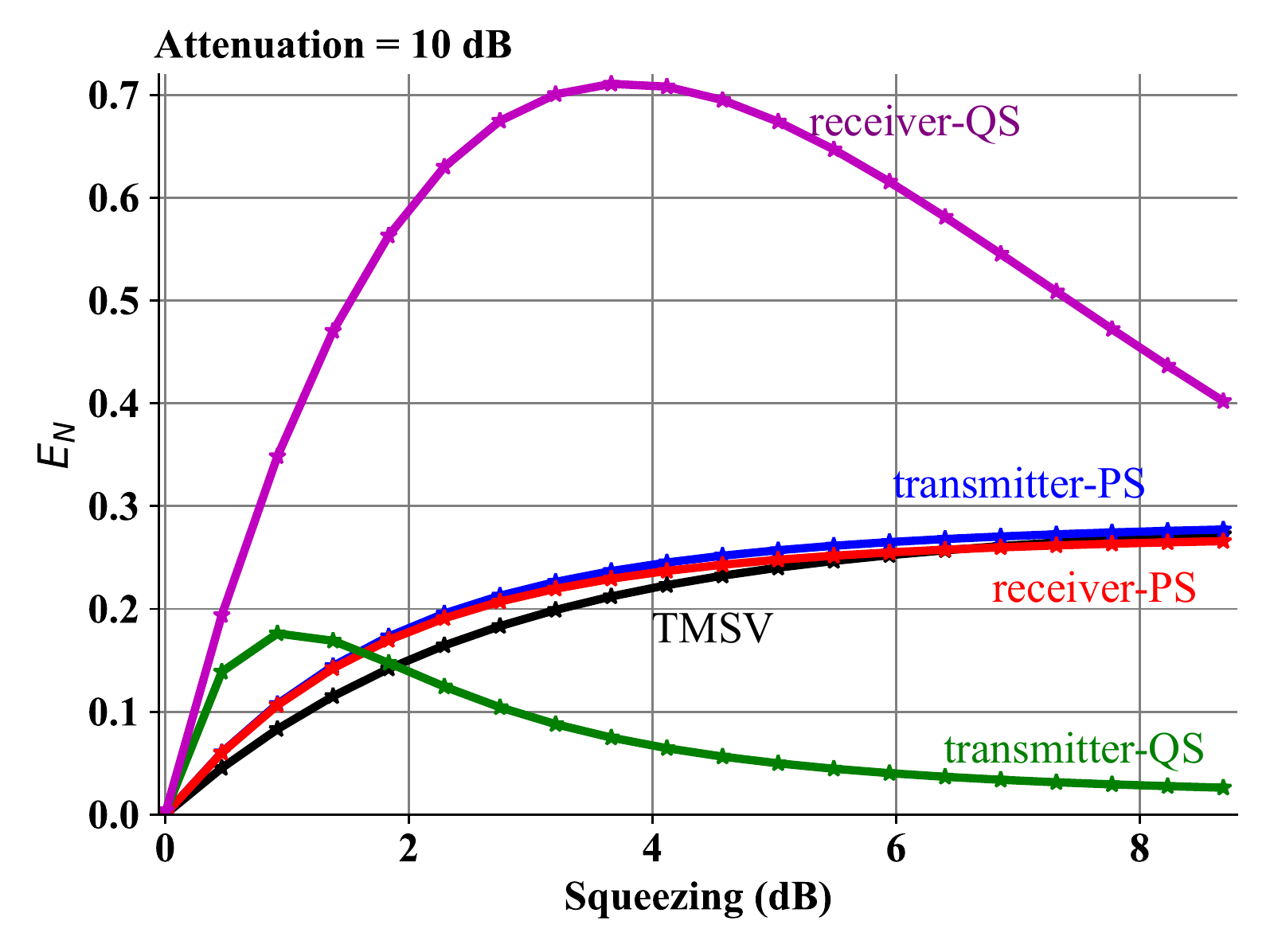}
\caption{The logarithmic negativity $E_N$ for PS and QS in the cases where each operation is applied successfully transmitter-side or receiver-side.
The parameters of both operations are optimized to maximize $E_N$, PS is optimized at a value of $\kappa_{PS}=0.95$ and $\kappa_{QS}=0.05$ for QS.}
\label{fig:EN}
\end{figure}

In the case with no photon loss, shown in Fig.\ref{fig:EN} (top), it makes no difference if the operations are applied receiver-side or transmitter-side, as both scenarios are equivalent.
We observe that for squeezing below 2dB QS outperforms PS in the amount of entanglement distilled.
Under 5dB and 10dB of photon loss, we see in Fig.\ref{fig:EN} (middle and bottom respectively) that receiver-QS becomes capable of increasing the entanglement at increasingly higher levels of squeezing. This fact is due to the photon loss causing the TMSV states to satisfy eq. \ref{eq:stateQS}. Conversely, this is not the case when the operation is applied transmitter-side, in this scenario we see that transmitter-QS reduces the entanglement of highly squeezed TMSV states.
On the other hand, we see that for every level of photon loss both receiver-PS and transmitter-PS are capable of increasing by a small margin the entanglement of the TMSV state.

\begin{figure}
\centering
\includegraphics[width=.40\textwidth]{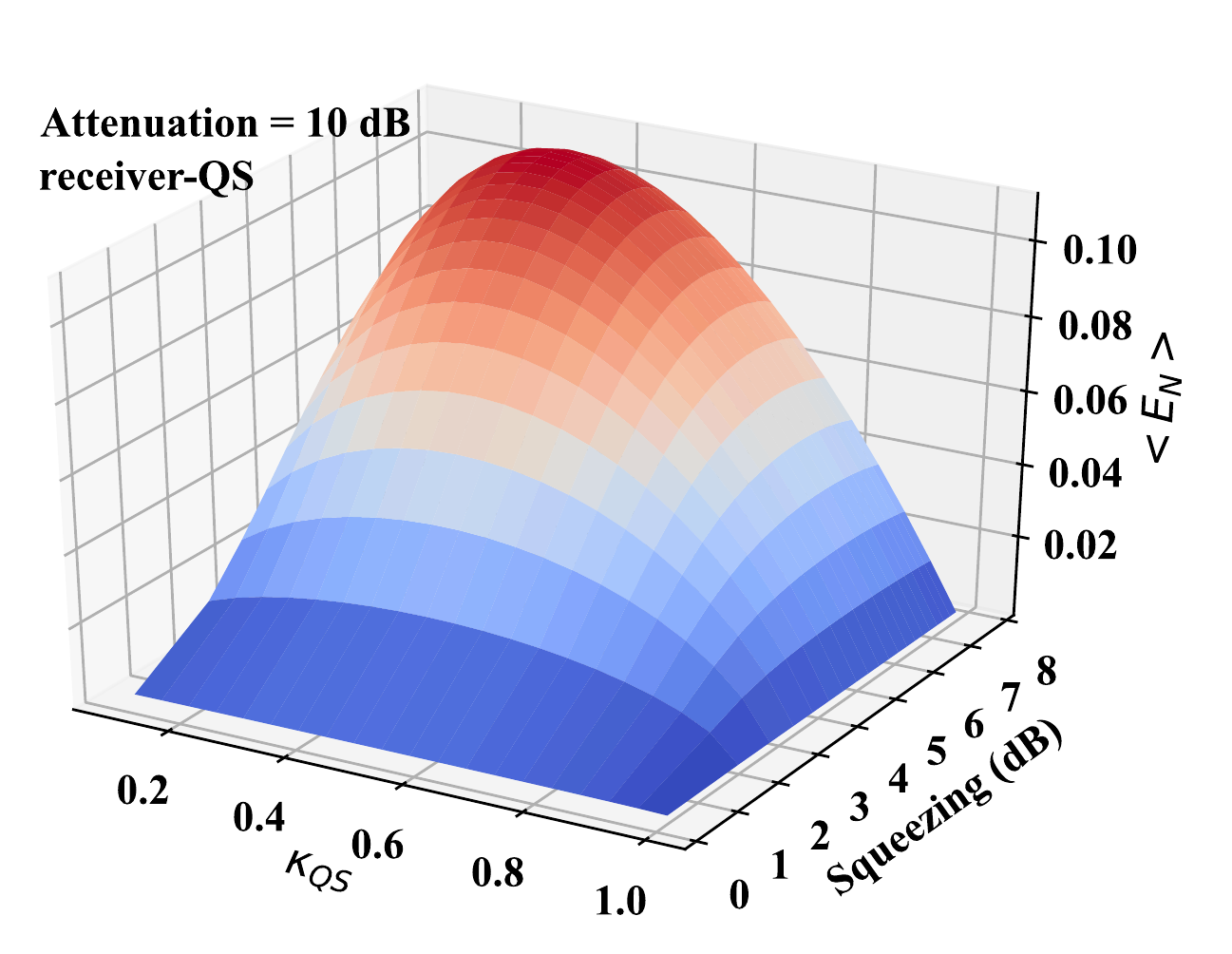}
\caption{The entanglement rate obtained after multiple applications of a receiver-QS, as a function of the initial squeezing of the TMSV and the parameter $\kappa_{QS}$.
The photon loss is fixed at 10 dB.}
\label{fig:EN_avg}
\end{figure}

Since both PS and QS are non-deterministic operations, we must also analyse the entanglement rate $\langle E_N \rangle$ achieved over multiple uses of the operations.
To compute this quantity we weight the values of $E_N$ obtained for each state by the probability of success of the corresponding operation.
In general, for the non-Gaussian operations we see that $\langle E_N \rangle$ is lower that $E_N$ for the TMSV.
In Fig.\ref{fig:EN_avg} we plot $\langle E_N \rangle$ for receiver-QS as a function of $\kappa_{QS}$ and the initial squeezing.
We can see that the maximum value of $\langle E_N \rangle$ that can be reached is $\approx 0.1$, compared with the value of $\approx 0.27$ (Fig.\ref{fig:EN} bottom) obtained for the TMSV state. Additionally, we repeated the calculation presented in Fig.\ref{fig:EN_avg} for both transmitter-PS and receiver-PS and we found $\langle E_N \rangle$ never exceeds the values of $E_N$ obtained from the TMSV state.
This fact is largely because optimal values $\kappa_{QS}$ and $\kappa_{PS}$ correspond to very low probabilities of success.

\subsection{Key rates}
\begin{figure}
\centering
\includegraphics[width=.41\textwidth]{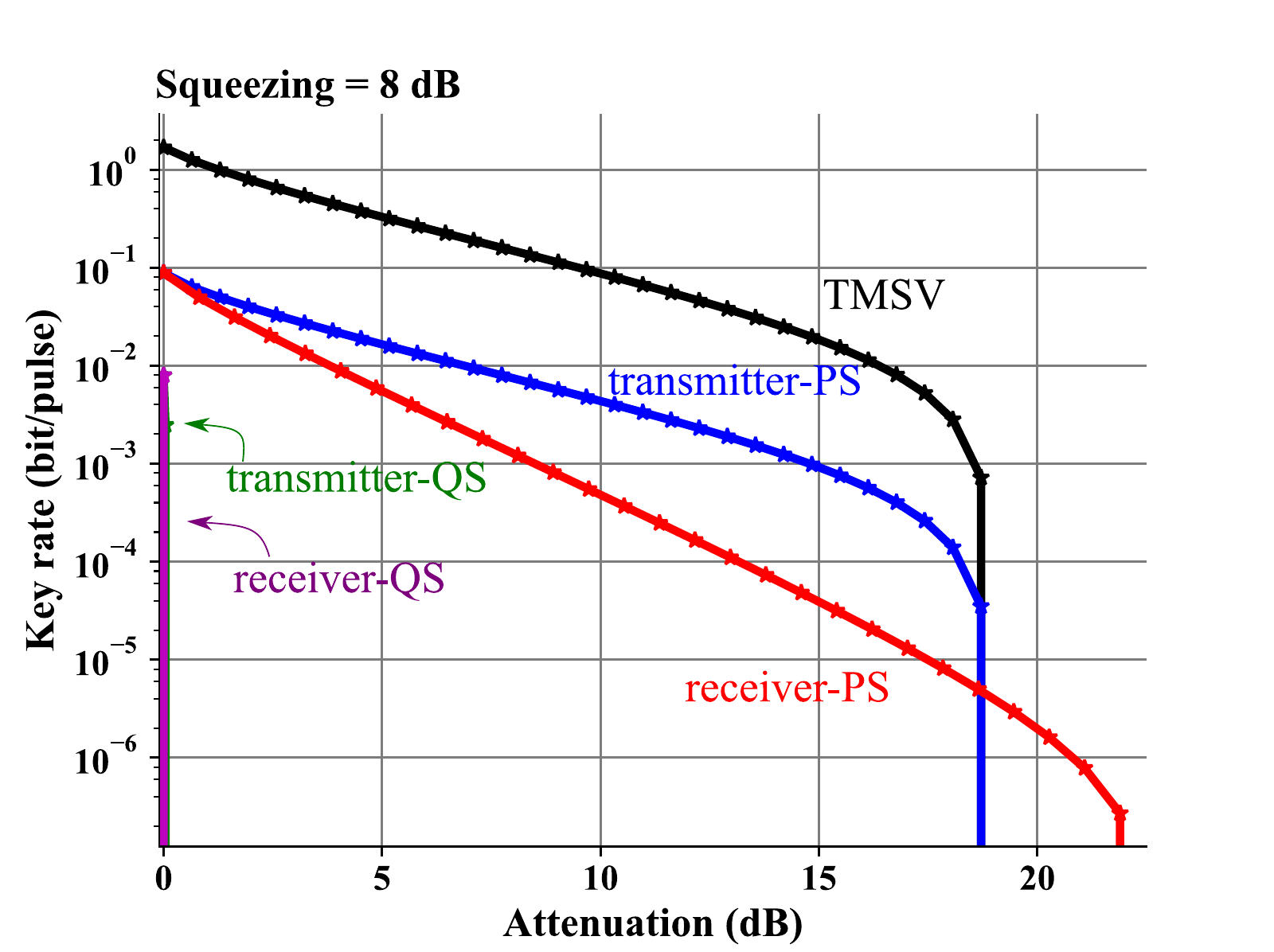}
\includegraphics[width=.41\textwidth]{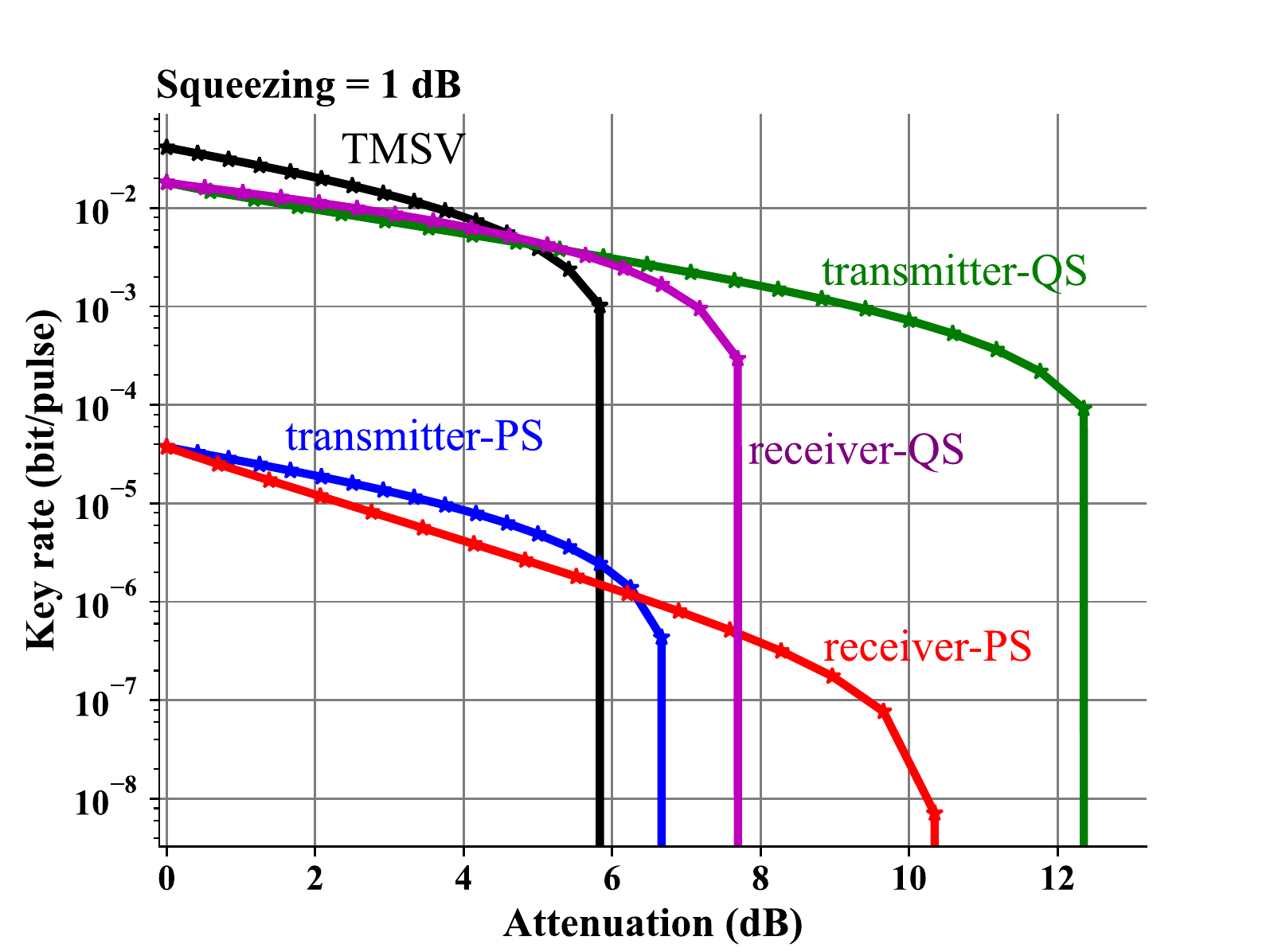}
\caption{Key rates for the operations as a function of photon loss.
The initial squeezing of the TMSV is set at 8dB and 1dB.}
\label{fig:KR}
\end{figure}

We present the values of the key rates obtained under the different operations.
In the case of 8dB of initial squeezing (Fig.\ref{fig:KR} top), using receiver-PS we obtained non-zero key rates for higher photon loss relative to the other operations. Nonetheless, the magnitude of the key rate of receiver-PS is lower than the one obtained by means of the TMSV state.
This result was originally presented in \cite{MingjianPS}.
For the squeezing value of 8dB we see that QS at both receiver-side and transmitter-side are not effective, being unable to recover any non-zero key rates for any non-zero photon loss.
On the other hand, we see that when the TMSV state has a low squeezing of 1dB (Fig.\ref{fig:KR} bottom) the best results where obtained by means of QS, with transmitter-QS producing the highest key rates between both transmitter-QS and receiver-QS. This is despite receiver-QS showing the greater increase of entanglement, as presented in Fig.\ref{fig:EN}. In this case receiver-QS also causes the entanglement between the mode $B$ and Eve's mode $E$ to increase. The increase in entanglement translates as an increase in Eve's Holevo information $\chi_{BE}$ involved in the calculation of the key rates.

\begin{figure}
\centering
\includegraphics[width=.48\textwidth]{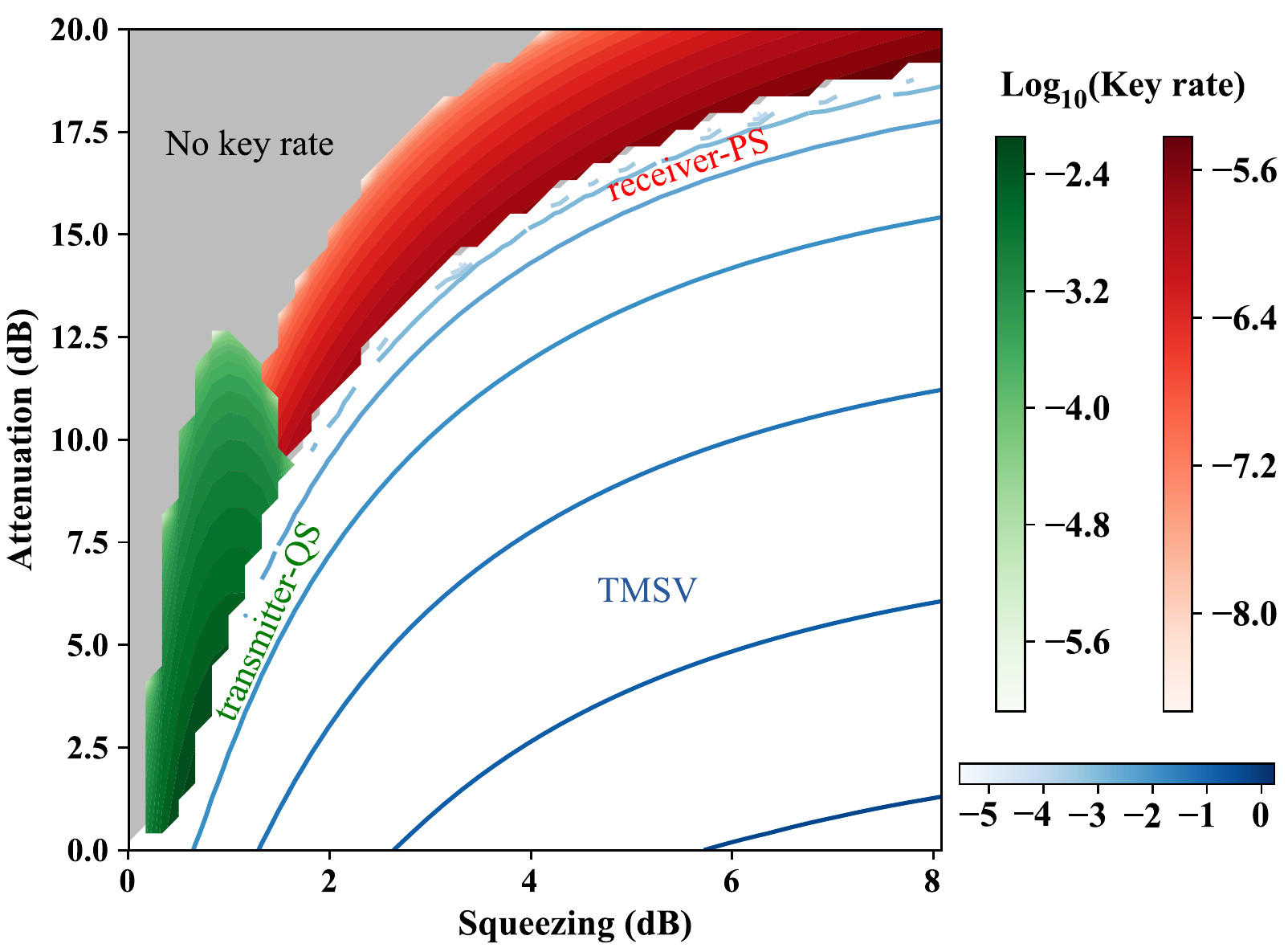}
\caption{Contour plots of $\log_{10}(\text{Key rate})$ as a function of photon loss and the initial squeezing of the TMSV states. The key rate is presented in units of bit/pulse.
The areas in green, red and blue represent the areas in which the highest key rates are achieved by transmitter-QS, receiver-PS and TMSV states respectively.
The area in grey corresponds to the parameters for which no key rates can be obtained by any operation.}
\label{fig:KRspace}
\end{figure}

To further investigate the key rates produced by transmitter-QS and receiver-PS, in Fig.\ref{fig:KRspace} we compare the key rates as function of both squeezing and photon loss.
Here we see the main result of this paper, the fact that for a given initial squeezing the non-Gaussian operations are capable of generating non-zero key rates at photon losses for which the Gaussian TMSV states are incapable of doing so.
Additionally,
we see for  initial squeezing levels below  1.5dB higher key rates are obtained by  transmitter-QS.
In these conditions transmitter-QS produces non-zero key rates for higher magnitudes of photon loss up to 12dB.
On the contrary, for values of squeezing beyond 1.5dB receiver-PS is capable of obtaining non-zero key rates for increasingly higher levels of photon loss, albeit at the cost of reduced key rate magnitudes.

As a side note, we point to the work presented in \cite{ScissorsQKD, ScissorsQKD2}. In  these works the authors make use of the QS operation in the receiver side to improve the key rates during the CV-QKD protocol introduced by Grosshans and Grangier \cite{GG02}.
The results presented in both works show that the QS enhanced QKD protocol can tolerate additional noise compared to the purely Gaussian QKD protocol.

\section{Conclusions}
We have studied the performance of non-Gaussian states in the entanglement based CV-QKD protocol. The non-Gaussian states considered were produced by means of applying the QS or PS operations to TMSV states. For each operation we considered scenarios where the operations are applied either transmitter-side or receiver-side.
In particular, we have calculated the  secure key rates produced by QKD as functions of photon loss and the initial squeezing of the TMSV states.

For PS, we observed that receiver-PS produced states that are more robust to loss compared to their Gaussian counterparts. Robust meaning that the states were capable of producing non-zero key rates for higher values of photon loss.
Conversely, we did not find any benefit in using transmitter-PS, since the states it produced resulted in lower secure key rates relative to those produced by the initial TMSV states.

For QS, we observed that the operation can increase the key rates obtained when used over TMSV states with initial squeezing below 2dB. Between receiver-QS and transmitter-QS, we observed that transmitter-QS yields the best results in terms of both the magnitude of the secure key rates and robustness against photon loss.

The main result of this work is that for a given initial squeezing of TMSV states,  the non-Gaussian operations, receiver-PS or transmitter-QS, increased the range of photon loss over which non-zero key rates could be obtained.
We found a threshold  of 1.5dB in the initial TMSV squeezing, above which receiver-PS becomes the preferred operation.
Alternatively, for an initial TMSV squeezing below 1.5dB, we found that transmitter-QS produces the highest secure key rates.
The results obtained here should help us understand how non-Gaussian operations can be deployed in Earth-satellite channels when maximisation of secure key rates is the goal.

\bibliographystyle{unsrt}

\bibliography{bib}

\end{document}